\def\aa{{A\&A}}

\def\aj{{AJ}}

\def\apj{{ApJ}}
\def\apjl{{ApJL}}

\def\mnras{{MNRAS}}

\def\pasp{{PASP}}
\newcommand{\ta}{{$T_{auto}$}\,}
\newcommand{\md}{{$T$--$\Sigma$\,}}
\newcommand{\mr}{{$T$--R\,}}
\newcommand{\cin}{{$C_{in}$\,}}

\documentclass[cup5b]{caps}
\usepackage{graphicx}
\usepackage{amssymb}
\usepackage{ociwsymp3}   

\HeadText{R. C. Nichol} 

\def\plottwo#1#2{\centering \leavevmode
\includegraphics[width=.45\columnwidth]{#1} \hfil
\includegraphics[width=.45\columnwidth]{#2}}

\begin{document}

\pagenumbering{arabic}

\author[]{R. C. Nichol\\Dept. of Physics, Carnegie Mellon University}

%
%

\chapter{Clusters of Galaxies in the SDSS}

\begin{abstract}

I review here past and present research on clusters and groups of
galaxies within the Sloan Digital Sky Survey (SDSS). I begin with a
short review of the SDSS and efforts to find clusters of galaxies
using both the photometric and spectroscopic SDSS data.  In
particular, I discuss the C4 algorithm which is designed to search for
clusters and groups within a 7--dimensional data--space, {\it i.e.},
simultaneous clustering in both color and space. Also, the C4 catalog
has a well quantified selection function based on mock SDSS galaxy
catalogs constructed from the Hubble Volume simulation. The C4 catalog
is $>90\%$ complete, with $<10\%$ contamination, for halos of ${\rm
M_{200}} > 10^{14}{\rm M_{\odot}}$ at $z<0.14$. Furthermore, the
observed summed $r$-band luminosity of C4 clusters is linearly related
to ${\rm M_{200}}$ with $<30\%$ scatter at any given halo mass. I also
briefly review the selection and observation of Luminous Red Galaxies
(LRGs) and demonstrate that these galaxies have a similar clustering
strength as clusters and groups of galaxies. I outline a new
collaboration planning to obtain redshifts for 10,000 LRGs at
$0.4<z<0.7$ using the SDSS photometric data and the AAT 2dF
instrument. Finally, I review the role of clusters and groups of
galaxies in the study of galaxy properties as a function of
environment. In particular, I discuss the ``SFR--Density'' and
``Morphology--Radius'' relations for the SDSS and note that both of
these relationships have a {\it critical density} (or ``break'') at a
projected local galaxy density of $\simeq 1\,h_{75}^2\,{\rm Mpc^{-2}}$
(or between $1$ to $2$ virial radii). One possible physical mechanism
to explain this observed critical density is the stripping of warm gas
from the halos of in--falling spiral galaxies, thus leading to a slow
strangulation of star--formation in these galaxies. This scenario is
consistent with the recent discovery (within the SDSS) of an excess of
``Passive'' or ``Anemic'' spiral galaxies located within the in--fall
regions of C4 clusters.

\end{abstract}

\section{Introduction}

As demonstrated by this conference -- {\it Clusters of Galaxies:
Probes of Cosmological Structure and Galaxy Evolution} --, clusters
and groups of galaxies have a long history as key tracers of the
large--scale structure in the Universe and as laboratories within
which to study the physics of galaxy evolution. At the conference,
their important role as cosmological probes was reviewed by several
speakers, including Simon White, Alan Dressler, Gus Oemler and
Guinevere Kauffmann. Therefore, I will not dwell on justifying the
importance of clusters to cosmological research, but simply direct the
reader to the reviews by these authors.

Instead, I provide below a brief overview of the Sloan Digital Sky
Survey (SDSS), followed by a discussion of the cluster--finding
algorithms (Section \ref{algorithms}) used within the SDSS
collaboration and present new scientific results obtained from studies
of galaxies as a function of environment (Section \ref{environ}).

\subsection{The Sloan Digital Sky Survey}

The Sloan Digital Sky Survey (SDSS; York et al. 2000; Stoughton et
al. 2002) is a joint multi--color ($u,g,r,i,z$) imaging and medium
resolution (R=1800; 3700$\aa$ to 9100$\aa$) spectroscopic survey of
the northern hemisphere using a dedicated 2.5 meter telescope located
at the Apache Point Observatory near Sunspot in New Mexico. During
well--defined (seeing $<1.7''$) photometric conditions, the SDSS
employs a mosaic camera of 54 CCD chips to image the sky via the
drift-scanning technique (see Gunn et al. 1998). These data are
reduced at Fermilab using a dedicated photometric analysis pipeline
(PHOTO; see Lupton et al. 2001) and object catalogs obtained for
spectroscopic target selection. During non--photometric conditions,
the SDSS performs multi--object spectroscopy using two bench
spectrographs attached to the SDSS telescope and fed with 640 optical
fibers.  The other ends of these fibers are plugged into a
pre--drilled aluminum plate that is bent to follow the 3--degree focal
surface of the SDSS telescope. The reader is referred to Smith et al.
(2002), Blanton et al. (2003) and Pier et al. (2003) for more details
about the SDSS.

In this way, the SDSS plans to image the northern sky in 5--passbands
as well as obtaining spectra for $\sim10^6$ objects.  In addition to
the large amount of the data being collected, the quality of the SDSS
data is high, which is important for many of the scientific goals of
the survey. For example, the SDSS has a dedicated Photometric
Telescope (PT; Hogg et al. 2001) which is designed to provide the SDSS
with a global photometric calibration of a few percent over the whole
imaging survey. Also, SDSS spectra are spectrophotometrically
calibrated.

The SDSS in now in production mode and has been collecting data for
several years. As of January 2003, the SDSS had obtained 4470 deg$^2$
of imaging data (not unique area) and had measured a half million
spectra. The SDSS has just announced its first official data release
(see {\tt http://www.sdss.org/dr1/}).

\section{SDSS Cluster Catalogs}
\label{algorithms}

One of the fundamental science goals of the SDSS was to create new
catalogs of clusters from both the imaging and spectroscopic data.  In
Table \ref{cats}, I present a brief overview of past and present
efforts within the SDSS collaboration as part of the SDSS Cluster
Working Group. In this table, I provide an appropriate reference if
available, a brief description of the cluster--finding algorithm and
the SDSS data being used to find clusters. Each of these algorithms
have different strengthens and science goals, and a detailed knowledge
of their selection functions is required before a fair comparison can
be carried out between these different catalogs. The reader is
referred to the review of Postman (2002) for more discussion of this
point.  However, Bahcall et al. (2003) has begun this process by
performing a detailed comparison of the clusters found by both the AMF
and maxBCG algorithms. The product of this work is a joint catalog of
799 clusters in the redshift range $0.05<z_{est}<0.3$ selected from
$\sim400 {\rm deg^2}$ of early SDSS commissioning data.

\begin{table}[t]
\begin{center}
\caption{Overview of past and present efforts to find clusters and
groups of galaxies within the SDSS collaboration. In the data column, ``I'' is for SDSS imaging data and ``S'' is for SDSS spectral data.}
\label{cats}
\begin{tabular}{r|lc|l}\hline\hline
  Name & Reference & Data& Description \\ \hline maxBCG & Annis et al.
& I & Model colors of BCG with z and\\ & & & look for E/S0 ridge-line
\\ AMF &Kim et al. (2002) & I & Matched Filter algorithm looking for\\
& & & over-densities in luminosity \& space \\ CE &Goto et al. (2002)
& I & Color-cuts, then uses Sextractor \\ & & & to find and de-blend
clusters \\ F-O-F & Berlind et al.& S & Friends-of-friends algorithm
\\ Groups & Lee et al. (2003)  & I & Mimics Hicksons Groups criteria\\ 
BH & Bahcall et al. (2003) & I & Merger of maxBCG \& AMF \\
C4 & Nichol et
al. (2001) & I \& S & Simultaneous clustering of\\ & & &
galaxies in color \& space\\ \hline \hline
\end{tabular}
\end{center}
\end{table}

As first outlined in Nichol (2001), there are main four challenges to
producing a robust optically--selected catalog of clusters. These are:

\begin{enumerate}
\item{To eliminate projection effects. This problem has plagued
previous catalogs of clusters and has been discussed by many authors
(see, for example, Lucey et al. 1983; Sutherland 1988; Nichol et al.
1992; Postman et al. 1992; Miller 2000);}
\item{A full understanding of the selection function. This has been
traditionally ignored for optical cluster catalogs (see Bramel et
al. 2000; Kochanek et al. 2003), but is critical
for all statistical analyzes using the catalog;}
\item{To provide a robust mass estimator. Traditionally, this has been
the Achilles' Heel of optical catalogs, as richness is a poor
indicator of mass; and,}
\item{To cover a large dynamic range in both redshift and mass}
\end{enumerate}

I review below one SDSS cluster catalog I am involved with, in
collaboration with Chris Miller at CMU, that now meets these four
challenges and is comparable in quality to the best X--ray catalogs
of clusters ({\it e.g.}, the REFLEX catalog of 
B{\" o}hringer et al. 2001).

\section{The C4 Algorithm}

\begin{figure}[tp]
\centering \includegraphics[width=9cm,angle=0]{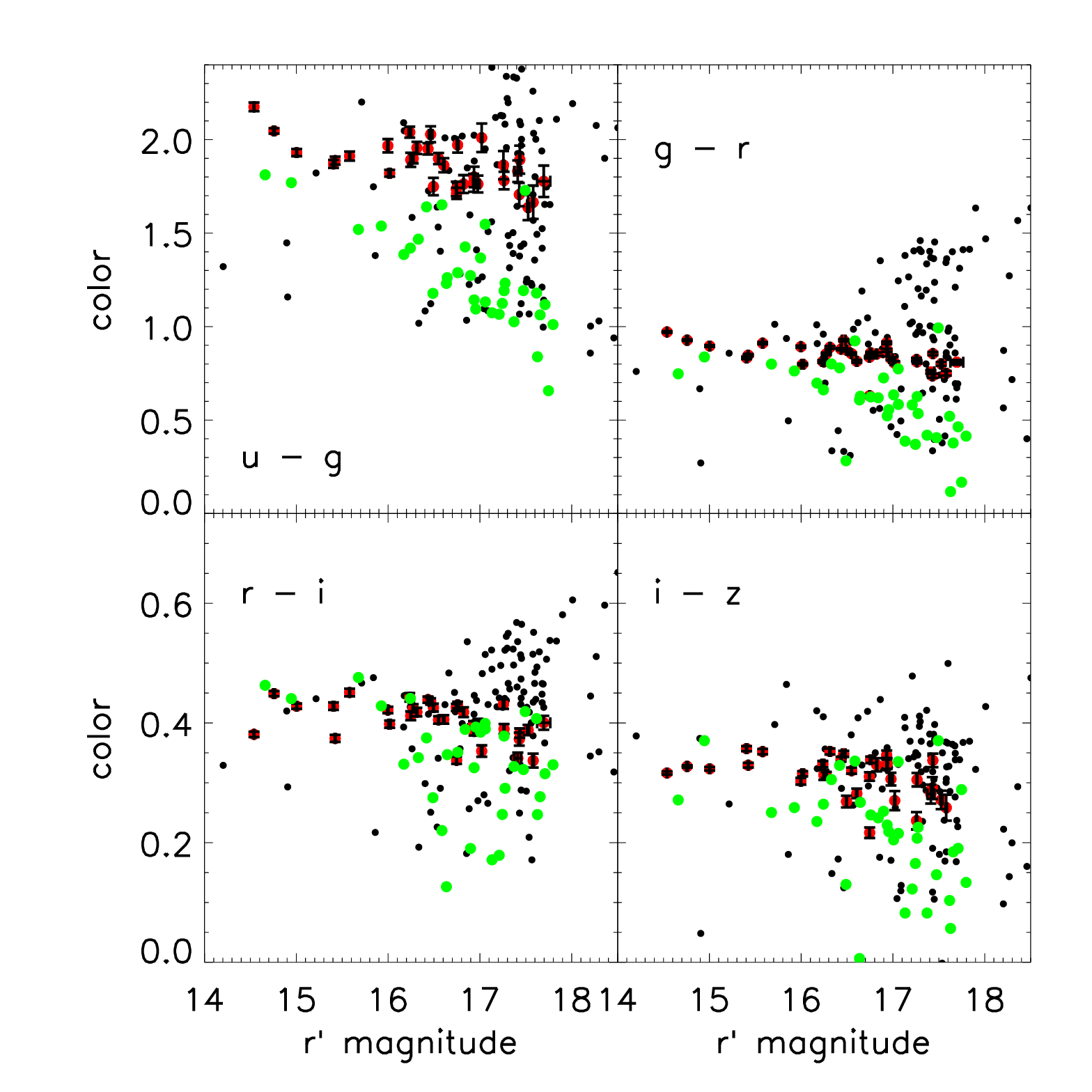}
\caption{The color-magnitude relations for a previously undiscovered
$z=0.06$ cluster in the Early Data Release of the SDSS detected by the
C4 algorithm.  The black dots are galaxies within an aperture of
$1h^{-1}$Mpc around the cluster center. The red and green dots are
actual cluster members (within $4\sigma_v$ in redshift space). The red
points are galaxies with a low $H_{\alpha}$ equivalent width, while
green points are galaxies with a high $H_{\alpha}$ equivalent width,
{\it i.e.}, passive and star--forming galaxies respectively.  The
errors on the colors are shown for the red points, indicating the
typical error bars on colors within the SDSS spectroscopic
sample. Note the tight correlation in color of the red points, which
is the E/S0 ridge line and is the signal the C4 algorithm is using to
find clusters.}
\label{cm}
\end{figure}

The underlying hypothesis of the C4 algorithm is that a cluster or
group of galaxies is a clustering of galaxies in both color and
space. This is demonstrated in Figure \ref{cm} and the reader is
referred to Gladders \& Yee (2000) for a full discuss of all the
evidence in support of this hypothesis.  Therefore, by searching for
clusters simultaneously in both color and space, the C4 algorithm
reduces projection effects to almost zero, while still retaining much
power for finding clusters (see Figure \ref{colorcolor}).

\subsection{Overview of C4 Algorithm}

I present here a brief overview of the C4 algorithm. To date, the C4
algorithm has been applied to the SDSS main galaxy spectroscopic
sample (Strauss et al. 2002) in the Early Data Release (EDR) of the
SDSS (see Stoughton et al. 2002; G{\' o}mez et al. 2003).

For each galaxy in the sample -- called the ``target'' galaxy below --
the C4 algorithm is performed in the following steps:

\begin{enumerate}

\item A 7--dimensional square box is placed on the target galaxy. The
center of this box is defined by the observed photometric colors ({\it
i.e.}, $u-g$, $g-r$, $r-i$, $i-z$), the Right Ascension, Declination
and redshift of the target galaxy. The width of the box is dependent
on the redshift and photometric errors on the colors of the target
galaxy. Once the box is defined, the number
of neighboring galaxies is counted within the SDSS main galaxy sample
(Strauss et al. 2002) inside this box and this count is reported (see
Figure \ref{colorcolor}).

\begin{figure}
\plottwo{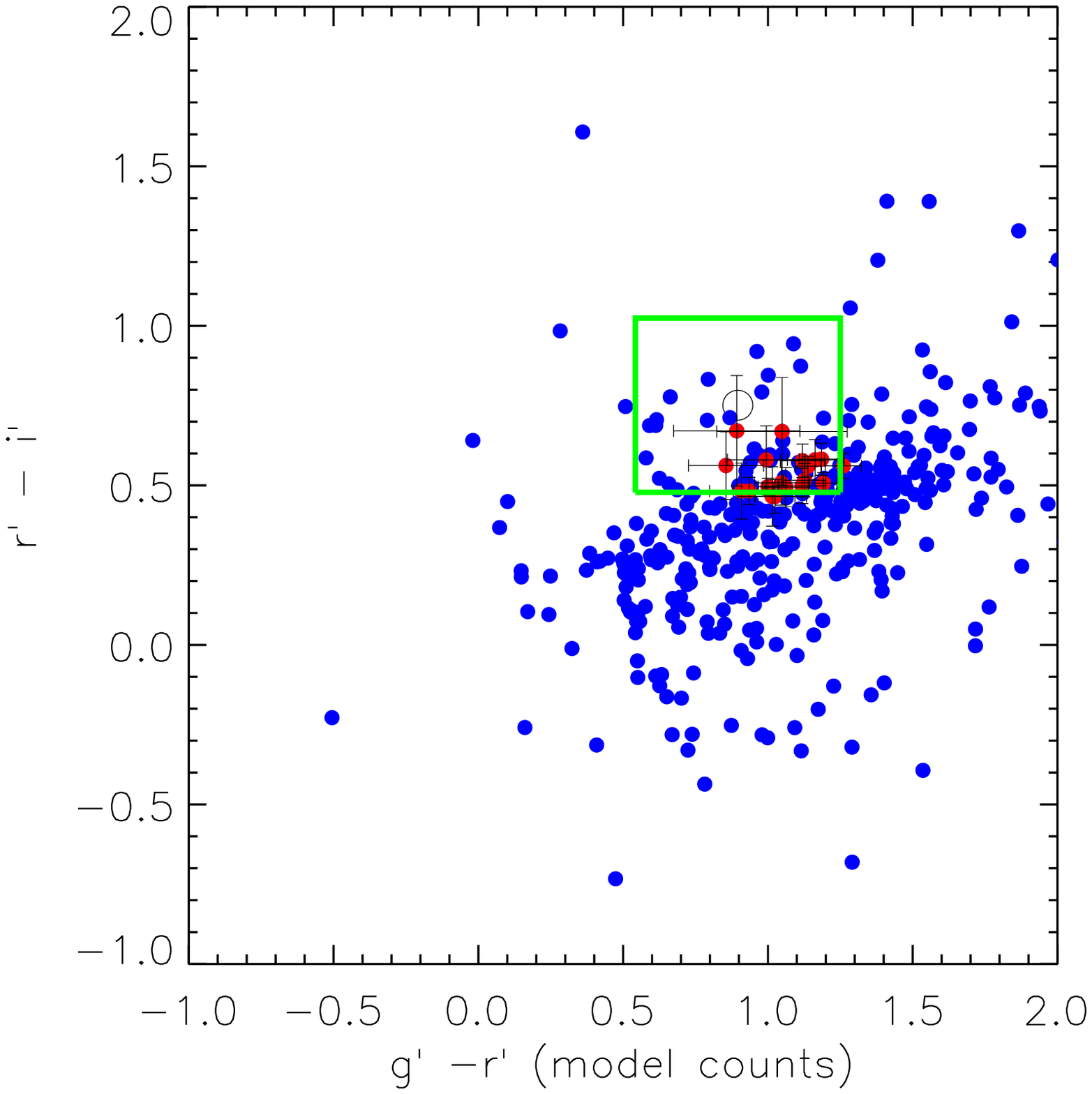}{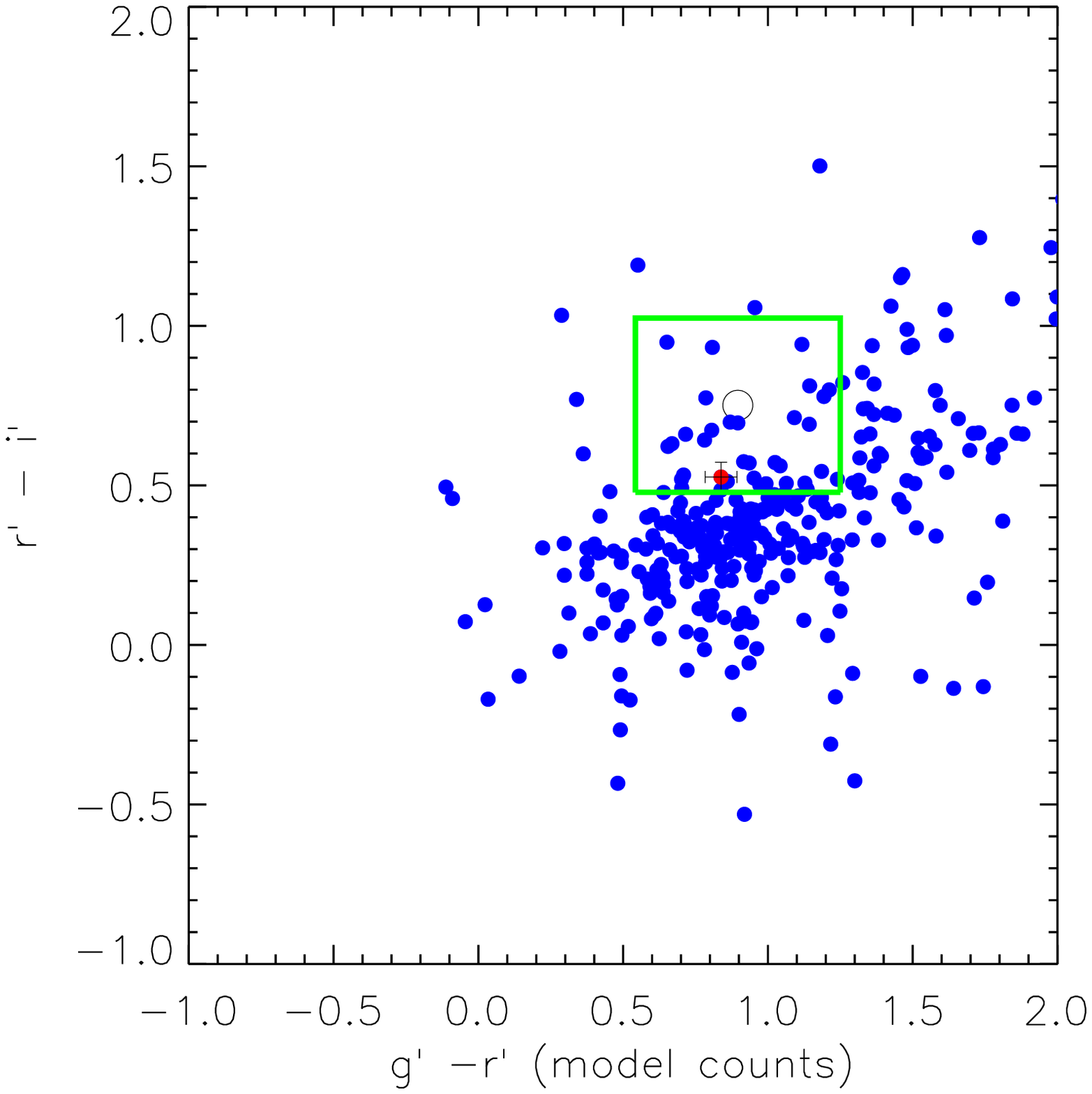}
\caption{Color-color plots for an example cluster galaxy (left) and
one randomly chosen field position (right). The blue dots are all
galaxies within the spatial part of the 7--dimensional search box
centered on the target galaxy ({\it i.e.}, all galaxies satisfying the
RA, Dec, and z dimensions of the 7D box). The red points are now those
galaxies which also lie within the color part of the 7--dimensional
search box, {\it i.e.}, they are close in both space and color. The
size of the 7D box in this color-color plane is shown in green. As you
can see it is much smaller than the scatter seen in the blue points
(close in just the spatial coordinates). For the random field position
(right), there is only one red point inside the box compared to ten
around the cluster galaxy (left). Therefore, all projection effects
have been eradicated as one does not expect false clustering in such a
high dimensional space.}
\label{colorcolor}
\end{figure}

\item The same 7--D box is placed on 100 randomly chosen galaxies,
also taken from the SDSS main galaxy spectroscopic sample, that
possess similar seeing and reddening values (see Figure
\ref{colorcolor}). For each of these 100 randomly chosen galaxies, the
number of neighboring galaxies is counted (from the SDSS main galaxy
sample) inside the box and a distribution of galaxy counts is
constructed from these 100 randomly chosen galaxies.

\item Using this observed distribution of galaxy counts, 
the probability of obtaining the observed galaxy count around the
original target galaxy is computed.

\item  This exercise is repeated for all galaxies in the sample and then
rank all these probabilities.

\item Using FDR (Miller et al. 2001) with an $\alpha=0.2$ ({\it i.e.},
only allowing a false discovery rate of 20\%), a threshold in
probability is determined that corresponds to galaxies that possess a
high count of nearest neighbors in their 7--dimensional box. 

\begin{figure}[tp]
\centering \includegraphics[width=10cm,angle=0]{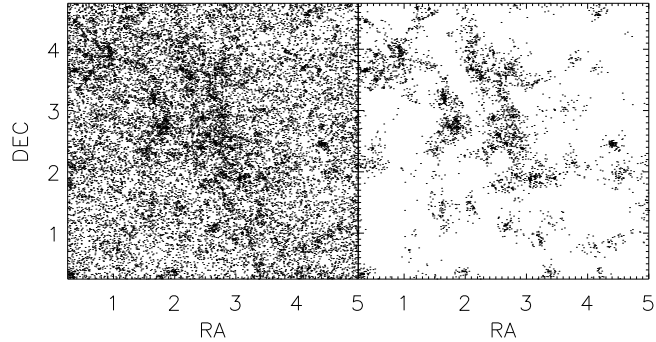}
\caption{The projected galaxy distribution in the simulations
before (left) and after (right) the C4 algorithm have been run and a
threshold applied to eliminate
field-like galaxies. This illustrates the effect of Step 6 in the algorithm.}
\label{before}
\end{figure}

\item All galaxies in our sample below this threshold in probability
are removed, which results in the eradication $\simeq80\%$ of all
galaxies. By design, the galaxies that are removed are preferentially
in low density regions of this 7--dimensional space ({\it i.e.}, the
field population). The galaxies that remain are called ``C4 galaxies''
which, by design, reside in high density regions with neighboring
galaxies that possess the same colors as the target galaxy. Figure
\ref{before} illustrates this step

\item Using only the C4 galaxies, the local density of all the C4
galaxies is determined using the distance to the $8^{th}$ nearest
neighbor.

\item The galaxies are rank order based on these measured densities
and then assigned to clusters based on this ranked list. This is the
same methodology as used to create halo catalogs within N--body
simulations (see Evrard et al. 2002).

\item This results in a list of clusters, for which a summed total
optical luminosity ($z$ and $r$ bands) and a velocity dispersion are
computed (see below).
 
\end{enumerate}

In summary, the C4 algorithm is a semi-parametric implementation of
adaptive kernel density estimation. The key difference with this
approach, compared to previous cluster--finding algorithms, is that it
does not model either the colors of cluster ellipticals ({\it e.g.},
Gladders \& Yee 2000, Goto et al. 2003), or the properties of the
clusters ({\it e.g.}, Postman et al. 1996; Kepner et al. 1999;  Kim et
al. 2002). Instead, the C4 algorithm only demands that the colors of
nearby galaxies are the same as the target galaxy. In this way, the C4
algorithm is sensitive to a diverse range of clusters and groups,
{\it e.g.}, it would detect a cluster dominated by a extremely
``blue'' population of galaxies (compared to the colors of field
galaxies). Therefore, the C4 catalog can be used for studying galaxy
evolution in clusters with little fear that the sample is biased
against certain types of systems.

\subsection{Simulations of the SDSS Data}

\begin{figure}[tp]
\centering \includegraphics[width=7cm,angle=0]{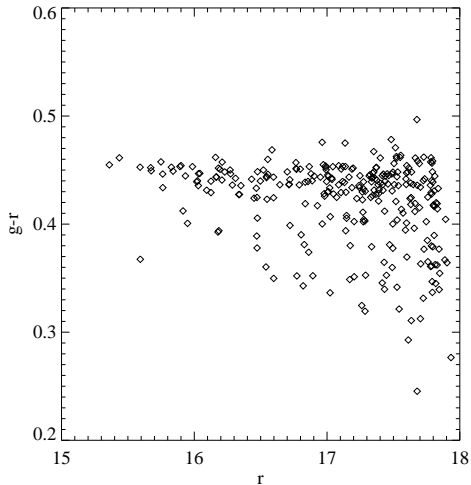}
\caption{The color--magnitude relation for a $10^{15}{\rm M_{\odot}}$
halo in the Hubble Volume simulation, where galaxies with SDSS colors 
and luminosities have been added to the simulation using the 
prescription outlined by Wechsler in this proceedings.}
\label{sim_cm}
\end{figure}

To address the four challenges given above, it is vital that we
construct realistic simulations of the C4 cluster catalog. This has
now been achieved through collaboration with Risa Wechsler, Gus Evrard
and Tim McKay at the University of Michigan.  Briefly, mock SDSS
galaxy catalogs have been created using the dark matter distribution
from the Hubble Volume simulations (Evrard et al 2002)\footnote{
http://www.mpa-garching.mpg.de/Virgo/hubble.html}.  The procedure for
populating the dark matter distribution with galaxies is described by
Wechsler in this proceedings, and in further detail by Wechsler et al
(in preparation), but I present a brief overview here.  The simulation
we use provides the dark matter distribution of the full sky out to
$z\sim 0.6$, where the dark matter clustering is computed on a light
cone.  Galaxies, with r-band luminosities following the luminosity
function found by Blanton et al (2001) for the SDSS, are added to the
simulation by choosing simulation particles (with mass
$\simeq2.2\times10^{12}\,{\rm M_{\odot}}$ per particle) from a
probability density function (PDF) of local mass density which depends
on galaxy luminosity.  This luminosity dependent PDF is tuned so that
the resulting galaxies match the luminosity dependent 2-point
correlation function (Zehavi et al. 2002).  Colors are then added to
the simulation using the colors of actual SDSS galaxies with similar
luminosities and local galaxy densities.  The resulting catalog thus
matches the luminosity function and color- and luminosity-dependent
2-point correlation function of the SDSS data.

Therefore, these mock SDSS catalogs produce clusters of galaxies
with a realistic color--magnitude diagram (compare Figures \ref{cm}
and \ref{sim_cm}). As we know the location of all the massive halos
({\it i.e.}, the clusters) within the Hubble Volume simulation (see
Jenkins et al. 2000 \& Evrard et al. 2002), we can now use these mock
catalogs to derive the selection function of the C4 algorithm. The
benefit of this approach is that these mock catalogs contain realistic
projection effects, as the galaxy clustering is constrained to match
the real data, and the clusters (halos) possess realistic profiles,
ellipticities and morphologies. This is a significant advancement on
past simulations used to quantify the selection function of cluster
catalogs (see Bramel et al. 2000; Postman et
al. 1996; Goto et al. 2002).

For comparison between the real and simulated data, two different
cluster observables are used. The first is the bi-weighted velocity
dispersion and the second is the summed optical luminosity of the
galaxies within the cluster.  To calculate the velocity dispersion
($\sigma_v$), an iterative technique is performed using the robust
bi-weighted statistics of Beers et al. (1990). The total optical
luminosity of the clusters is determined by converting the apparent
magnitudes of galaxies in the cluster to optical luminosities using
the conversions in Fukugita et al. (1996). All magnitudes are also
$k$-corrected according to Blanton et al. (2002) and extinction
corrected according to Schlegel, Finkbeiner, and Davis (1998).
Cluster membership is defined to be any galaxy within $4\sigma_v$ in
redshift, and within $1.5h^{-1}Mpc$ projected separation on the sky.
For the halo catalog, the mass within 200 times the critical density
($M_{200}$) is used as determined from summing up all the dark matter
particles within this radius around each halo ($R_{200}$; see Evrard
et al. 2002).

\subsection{Purity and Completeness: Challenges One and Two}
\label{purity}

\begin{figure}[tp]
\plottwo{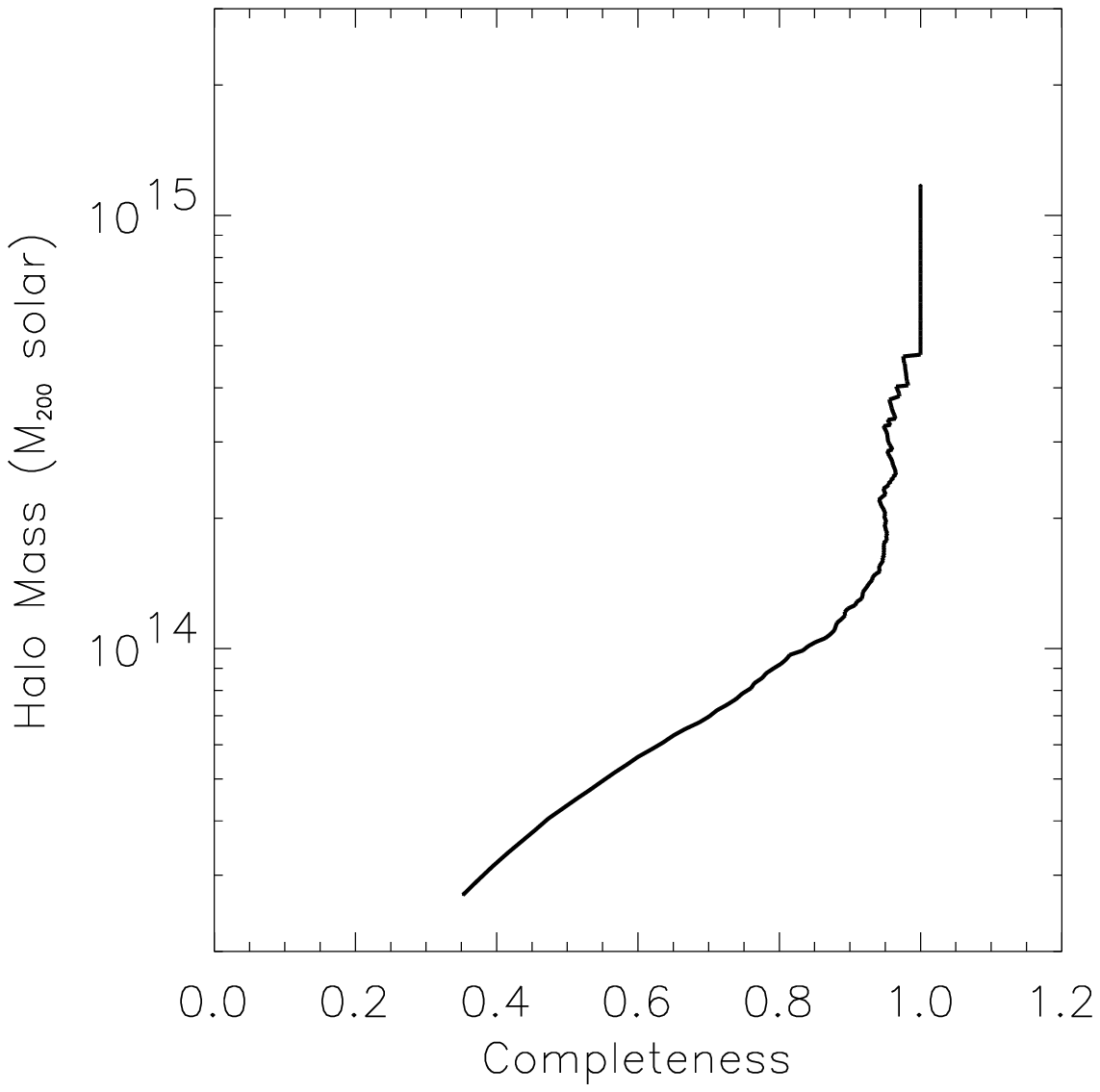}{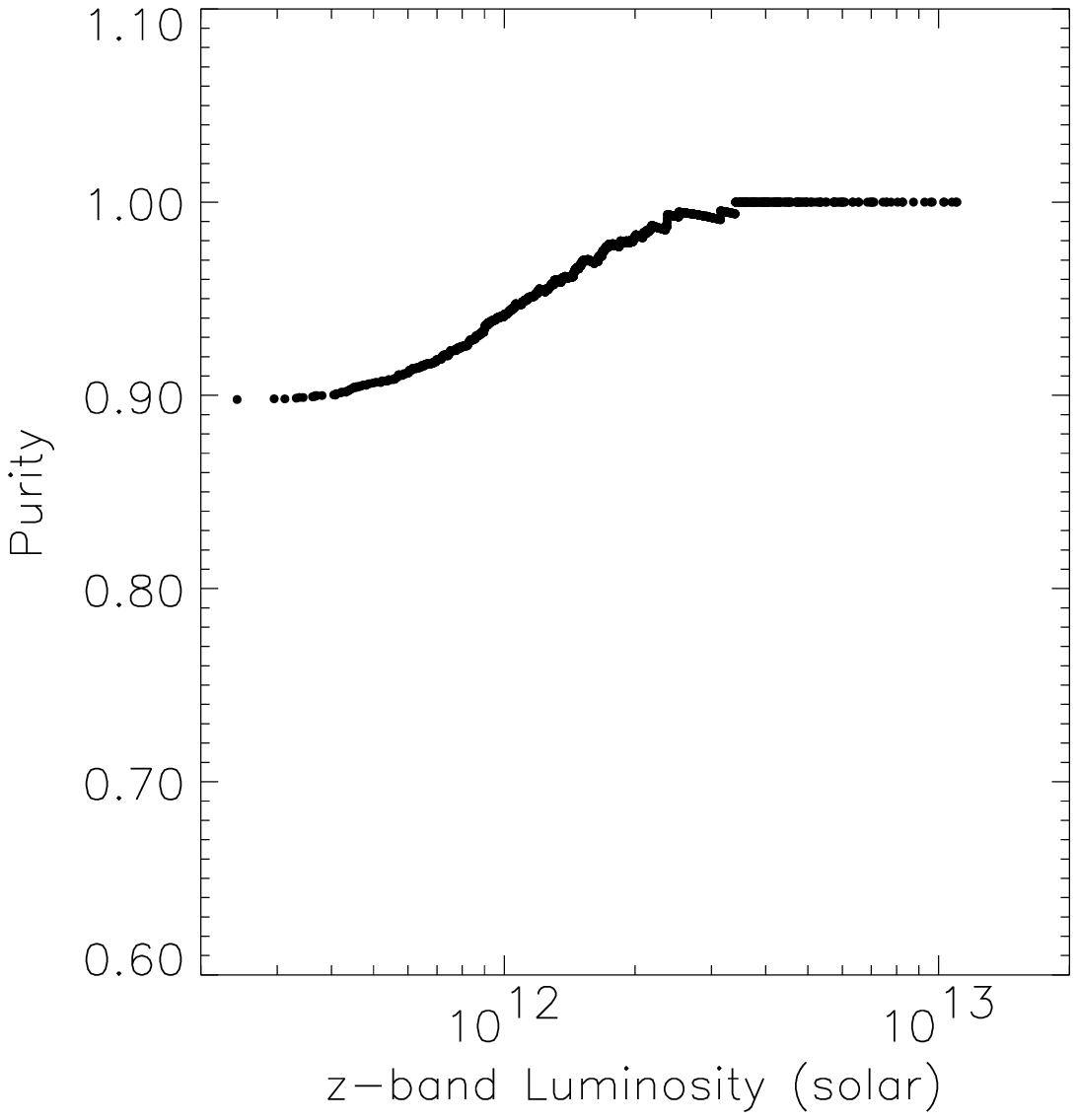}
\caption{(Left) The completeness of the C4 catalog as a
function of the dark matter halo mass from the simulation (${\rm
M_{200}}$). (Right) The purity of the C4 catalog, as a function of the
summed total $r$-band luminosity of the cluster, as derived from our
mock SDSS galaxy catalog.}
\label{completeness}
\end{figure}

The C4 algorithm has been applied to these mock SDSS galaxy catalogs.
In Figure \ref{completeness}, I show the purity of the matched C4
clusters found in the simulation as a function of the total $r$ band
luminosity. Purity is defined to be the percentage of systems detected
in the mock SDSS catalog, using the C4 algorithm, that match a known
dark matter halo in the HV simulation. Clearly, the purity of the C4
catalog remains high over nearly 2 orders of magnitude in luminosity
and is a direct result of searching for clustering in a
high--dimensional space where projection effects are rare.

In Figure \ref{completeness}, I also present the completeness of the
C4 catalog, as a function of the dark halo mass (${\rm M_{200}}$), and
demonstrate that the C4 catalog remains $>90\%$ complete for systems
with ${\rm M_{200}} > 10^{14} {\rm M_{\odot}}$. The accuracy of these
completeness measurements greatly benefits from the large sample sizes
available from the HV simulation. However, for the lower mass systems,
there may be some incompletenesses in the original dark matter halo
catalog because of the mass resolution of the HV simulations.
Furthermore, the purity and completeness of the C4 catalog is only a
weak function of the input parameters, {\it e.g.}, the FDR threshold,
and thus robust against the exact parameter choices.  In summary,
using these mock SDSS catalogs, we have addressed the first two
challenges given in Section \ref{algorithms}, {\it i.e.}, for
$z<0.14$, the C4 catalog is $>90\%$ complete, with $<10\%$
contamination, for halos of ${\rm M_{200}} > 10^{14}{\rm M_{\odot}}$.

\subsection{Mass Estimator and Dynamic Range: Challenges Three and Four}
\label{mass}

\begin{figure}[tp]
\centering \includegraphics[width=7cm,angle=0]{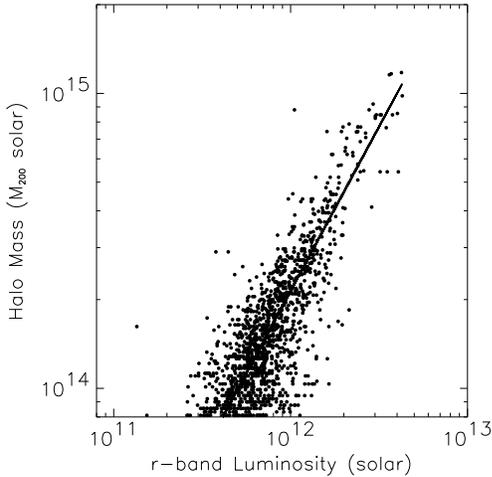}
\caption{The relation between the dark matter halo mass in our simulation
and the total summed $r$ luminosity. The line is the best fit and
the scatter, for a given mass, is $<30$\%.}
\label{matchesLr}
\end{figure}

In Figure \ref{matchesLr}, I present the correlation between the ${\rm
M_{200}}$ dark matter halo mass (taken directly from Evrard et
al. 2002) and the total summed $r$ band luminosities for 4734
clusters that match in the HV simulation. As expected, these two
quantities are linearly correlated over 2 orders of magnitude in both
the halo mass and optical luminosity.  The measured scatter for the
whole dataset is $25\%$ (measured perpendicular to the best fit
line). This plot demonstrates that we have address the 3$^{rd}$ and
4$^{th}$ challenges in Section \ref{algorithms}.

In the future, the C4 algorithm will be extended to include: {\it i)}
Higher resolution dark matter simulations to test the sensitivity of
the algorithm to lower mass systems; {\it ii)} A range of cosmological
simulations to probe our sensitivity to various cosmological
parameters {\it e.g.}, $\sigma_8$; and {\it iii)} Improve our methods of
populating the simulations with SDSS galaxies using higher order
statistics ({\it e.g.}, the 3--point correlation function).

\section{Luminous Red Galaxies}


I briefly review here forthcoming surveys of Luminous Red Galaxies
(LRGs) as these surveys will soon replace clusters as the most
efficient tracers of the large--scale structure in the Universe.  Such
galaxies are selected to be dominated by an old stellar population
(using the SDSS colors) and are luminous, so they can be seen to high
redshift even in the SDSS photometric data (see Eisenstein et al. 2001
for details).

A preliminary analysis of the correlation functions for
both a sample of LRGs, selected from the SDSS (Eisenstein et
al. 2001), and normal SDSS galaxies (Zehavi et al. 2001), demonstrates
that, as expected, the LRGs are more strongly clustered than normal
galaxies. The LRG correlation function has an amplitude and
scale--length consistent with that measured for groups and
clusters of galaxies ({\it e.g.}, Peacock \& Nicholson 1991; Nichol et
al. 1992; Collins et al. 2000; Miller 2000; Nichol 2002). Therefore, by design, a
large fraction of these LRGs must lie in dense environments as their
spatial distribution clearly traces the distribution of clusters and
groups in the universe. The main advantage of the LRG selection is
that it does not depend upon the details of finding clusters of
galaxies, and therefore, their selection is more straightforward to
model (see Eisenstein et al. 2001).

\begin{figure}[tp]
\centering \includegraphics[width=9cm,angle=0]{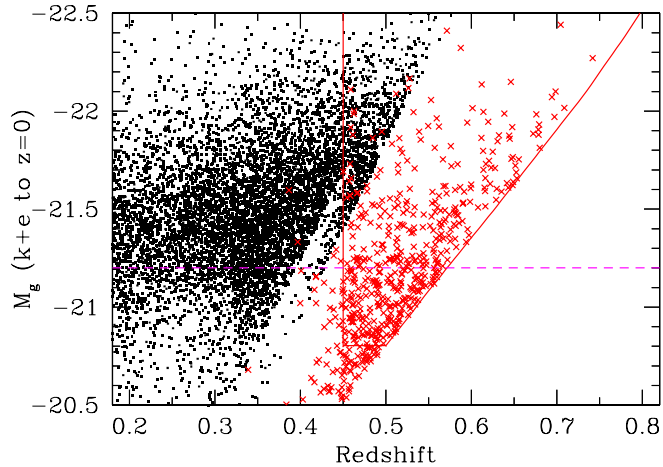}
\caption{The distribution in g--band luminosity and redshift for both
the low redshift LRGs (solid points) and the SDSS--2dF LRGs (red
crosses). The luminosities are k--corrected and corrected for passive
evolution to $z=0$. The red lines show the expected selection
boundaries for the SDSS-2dF survey, while the dashed line is the
luminosity limit above which we expect the low redshift SDSS LRG
sample to be complete. Plot courtesy of Daniel Eisenstein and my
SDSS--2dF colleagues.}
\label{2df}
\end{figure}

Due to the 45 minute spectroscopic exposure time of the SDSS, the SDSS
only targets LRG candidates brighter than r$=19.5$ (Eisenstein et
al. 2001). This corresponds to a cut--off at $z\simeq 0.45$, but one
can easily detect LRG candidates in the SDSS photometric data to
higher redshifts$\footnote{Beyond z=0.45, the SDSS LRG selection
becomes easier than at lower redshift (with less contamination)
because of the fortuitous design of the SDSS filter system, see
Eisenstein et al. (2001)}$. Therefore, using the unique 2dF
multi-object spectrograph on the Anglo-Australian Telescope, we have
begun a joint SDSS--2dF program to push the original SDSS LRG
selection to higher redshift. At the time of writing, our initial
observations have been very successful, with spectra for $\simeq1000$
LRGs in the redshift range of $0.4 < z < 0.7$. In Figure \ref{2df}, I
show the distribution in luminosity and redshift of these new
SDSS--2dF LRGs and highlight that they cover a comparable range in
their luminosities as the low redshift LRGs. By the end of the
SDSS--2dF LRG survey, we hope to have redshifts for 10,000 LRGs over
this intermediate redshifts range. When combined with the low redshift
SDSS LRGs, we will be able to study the evolution in the properties
and clustering of a single population of massive galaxies over half
the age of the universe.

\section{Galaxy Properties as a Function of Environment}
\label{environ}

Clusters and groups of galaxies play an important role in studying the
effects of environment on the properties of galaxies.  With the SDSS
data, it is now possible to extend such studies well beyond the cores
of clusters into lower density environments. Furthermore, the distance
to the $N^{th}$ nearest neighbor can be used to provides an adaptive
measure of the local density of galaxies (see Dressler et al. 1980;
Lewis et al. 2002; G{\' o}mez et al. 2003). One can also use kernel density
estimators ({\it e.g.}, Eisenstein 2003) and mark correlation
functions.

There were many great talks and posters on the topic of galaxy
evolution in clusters at this conference. For example, see the
contributions by Bower, Davis, Dressler, Franx, Goto, Koo, Kauffmann,
Martini, N. \& C. Miller, Postman, Poggianti, Tran and Treu. Also, I
refer the reader to the work of Hogg et al. (2003) who is also using
the SDSS data to study the effects of environment on the colors,
surface brightnesses, morphologies and luminosities of galaxies.

\subsection{Critical Density}

\begin{figure}[tp]
\plottwo{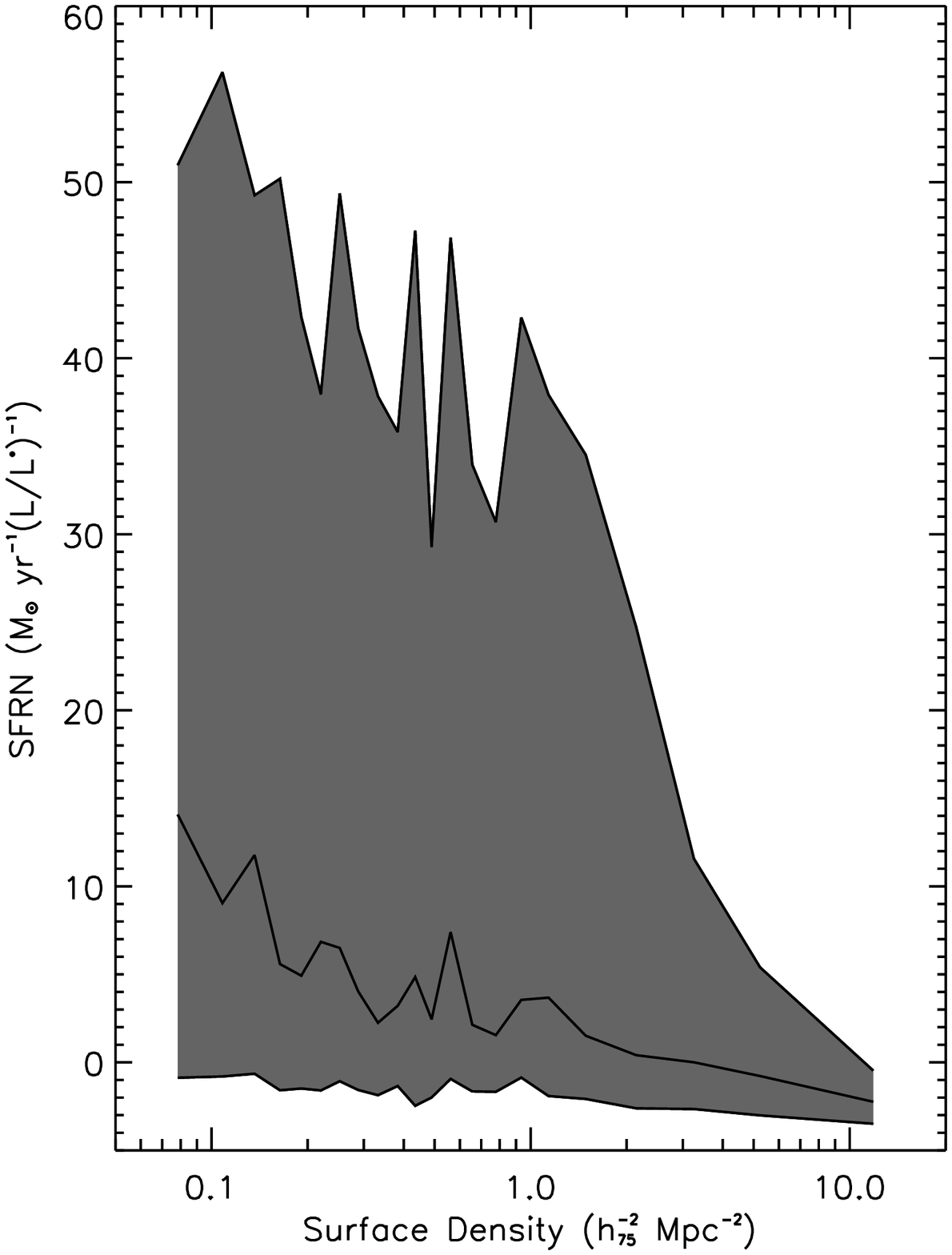}{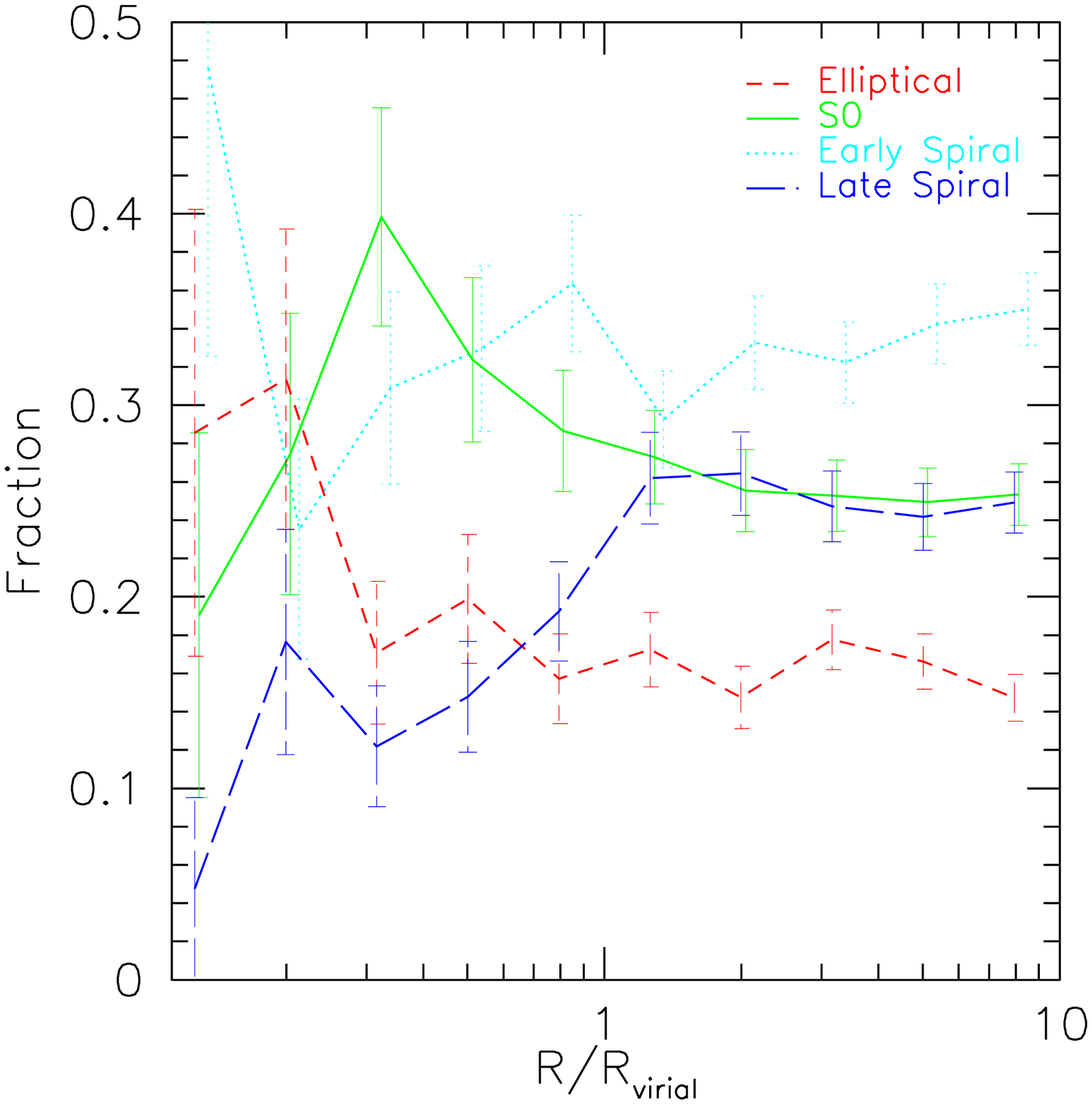}
\caption{(Left) The relation between SFR and density for a
volume--limited sample of 8598 SDSS galaxies between
$0.05<z<0.095$. The y--axis is the normalized SFR, {\it i.e.}, the SFR
divided by the $z$ luminosity of the galaxy. The x--axis is the
projected local surface density computed from the distance to the
$10^{th}$ nearest neighbor in a redshift shell of $\pm1000{\rm km\,
s^{-1}}$. See G{\' o}mez et al. (2003) for details. The top of the shaded
region is the $75^{th}$ percentile of the distribution, while the
bottom is the $25^{th}$ percentile.  The solid line through the shaded
area is the median of the distribution. As one can see, beyond a
density of $\simeq 1\,h_{75}^2\,{\rm Mpc^{-2}}$, the tail of the distribution
($75^{th}$ percentile) is heavily curtailed in dense
environments. (Right) The morphology--radius relation for C4 clusters
in the SDSS. The morphologies were derived using the \ta parameter
discussed in the text, while the distance to the nearest C4 cluster
(y--axis) has been scaled by the virial radius of that clusters. These
plots were taken from G{\' o}mez et al. (2003) and Goto et al. (in prep).}
\label{critical}
\end{figure}

Is the star--formation rate (SFR) of a galaxy affected by its
environment? The answer appears to be yes, and was discussed by
several authors at this conference (see the contributions by Miller,
Couch and Bower). In particular, G{\' o}mez et al. (2003) find that the
fraction of strongly star--forming galaxies in the SDSS decreases
rapidly beyond a critical density of $\simeq 1\,h_{75}^2\,{\rm
Mpc^{-2}}$ (see also Lewis et al. 2002). This result is demonstrated
in Figure \ref{critical} and appears to be the same for all
morphological types (see G{\' o}mez et al. 2003 and below).

In Figure \ref{critical}, we also show a preliminary SDSS {\it
Morphology--Radius} (\mr) relation based on the work of Goto et
al. (in prep). The morphological classifications used in this figure
are based on new, and improved, concentration index ($C_{in}$; see
Shimasaku et al. 2001) measurement of Yamauchi et
al. (in prep). Briefly, Yamauchi et al. (in prep) compute their
concentration index within two--dimensional, elliptical isotopes,
which account for the observed ellipticity and orientation of the
galaxy on the sky. This improvement helps prevent low inclination
galaxies ({\it e.g.}, edge--on spirals) from being mis--classified as
early-type galaxies. Furthermore, Yamauchi et al. (in prep) computes the
``coarseness'' of each galaxy, which is a measure of the residual
variance after the best fit 2--D galaxy model has been
subtracted. This coarseness measurement can therefore detect the
presense of spiral arms in a galaxy. These two measurements of the
morphology are re--normalized (by their rms) and added
to produce a final morphological parameter called \ta. Yamauchi et
al. (in prep) has tested their algorithm extensively and have
demonstrated that \ta is more strongly correlated with visual
morphological classifications than the normal \cin parameter, with a
correlation coefficient with the visual morphologies of 0.89.



Using the \ta classification, Goto et al. (in prep) has separated
SDSS galaxies into the four (traditional) morphological subsamples of
ellipticals, lenticulars (S0's), early spirals (Sa, Sb galaxies) and
late spirals (Sc and Irregulars). This is presented in Figure
\ref{critical}. Clearly, the mapping between the \ta parameter and
these visually--derived morphological classifications is not perfect,
but such an analysis does allow for an easier comparison with previous
measurements of the \mr and Morphology--Density (\md) relations, and
theoretical predictions of these relationships (see Benson et
al. 2000).


The SDSS \mr relation shown in Figure \ref{critical} remains constant
at $>2$ virial radii from C4 clusters. As expected, this corresponds
to low density regions ($<1\,h_{75}^2\,{\rm Mpc^{-2}}$) in our
volume--limited sample. This observation is consistent with previous
determinations of the \md and \mr relations in that these functions
are near constant at low densities (see Dressler et al 1980; Postman
et al. 1984; Dressler et al. 1997; Treu et al. 2003). At a radius of
$\simeq1$ virial radius, we witness a change in the \mr relation, {\it
i.e.}, we see a decrease in the fraction of late spiral galaxies
($\sim$Sc galaxies) with smaller cluster-centric radii. We also see
some indication of a decrease in the early spirals. It is interesting
to note that the critical density of $\simeq 1\,h_{75}^2\,{\rm
Mpc^{-2}}$ seen in the SFR--Density relation of G{\' o}mez et
al. corresponds to a cluster--centric radius of between $\simeq 1$ to
$2$ virial radii. The key question is; Are these two phenomena just
different manifestations of the same physical process which is
transforming both the morphology and SFR of the galaxies at this
critical density? I believe the jury is still out on this question
(but see the contributions of Bower and Miller in these proceedings).

I note here that Postman \& Geller (1984) also reported a critical
density (or ``break'') in their \md relation at approximately the same
density as seen in Figure \ref{critical}, {\it i.e.}, $\simeq 3.5
h^3_{75}{\rm Mpc^{-3}}$. Therefore, this ``break'' (or critical
density) seen in the \md relation appears to be universal, as it has
been seen in two separate studies, which are based on different
selection criteria and analysis techniques.

\begin{figure}[tp]
\plottwo{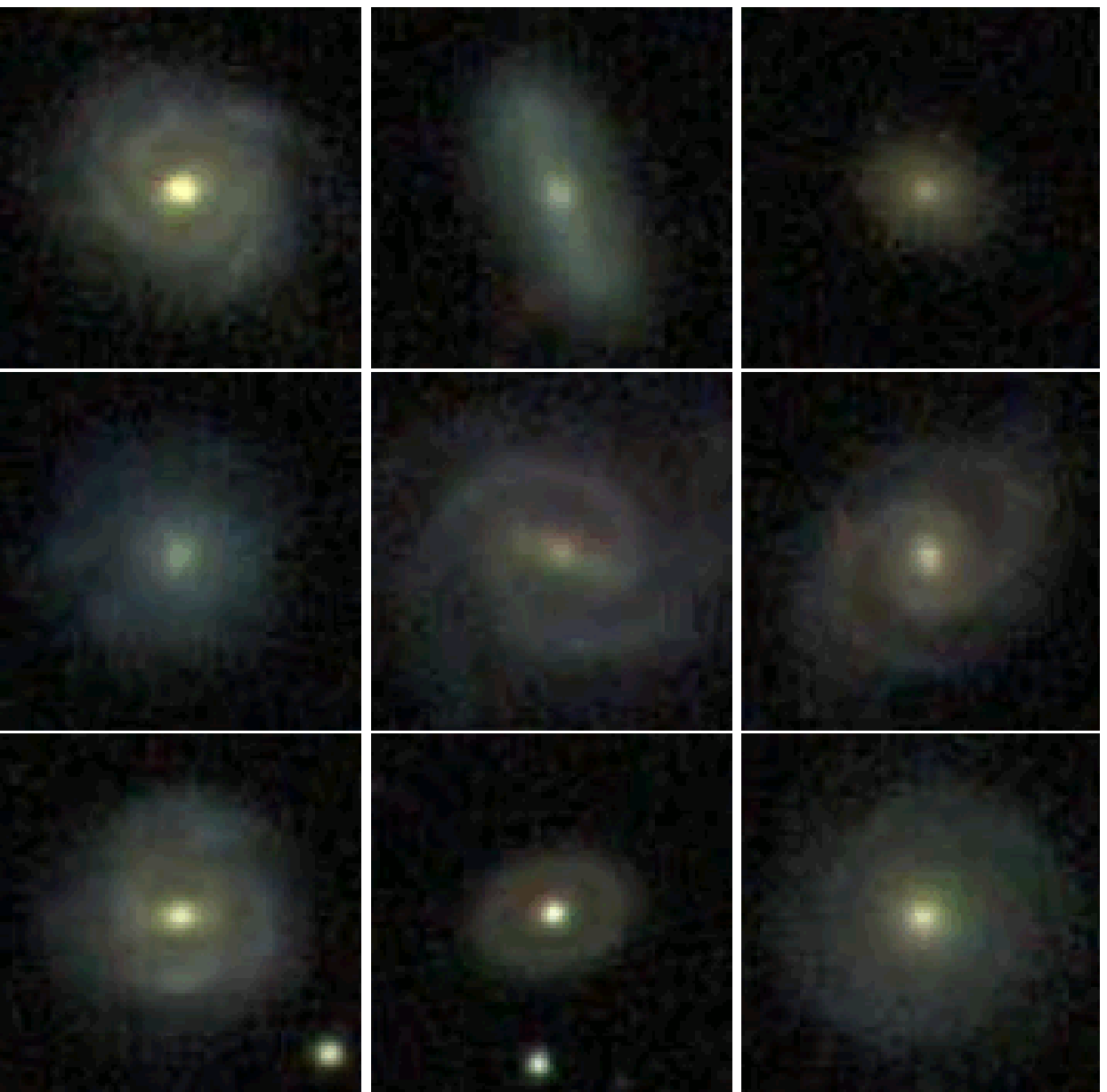}{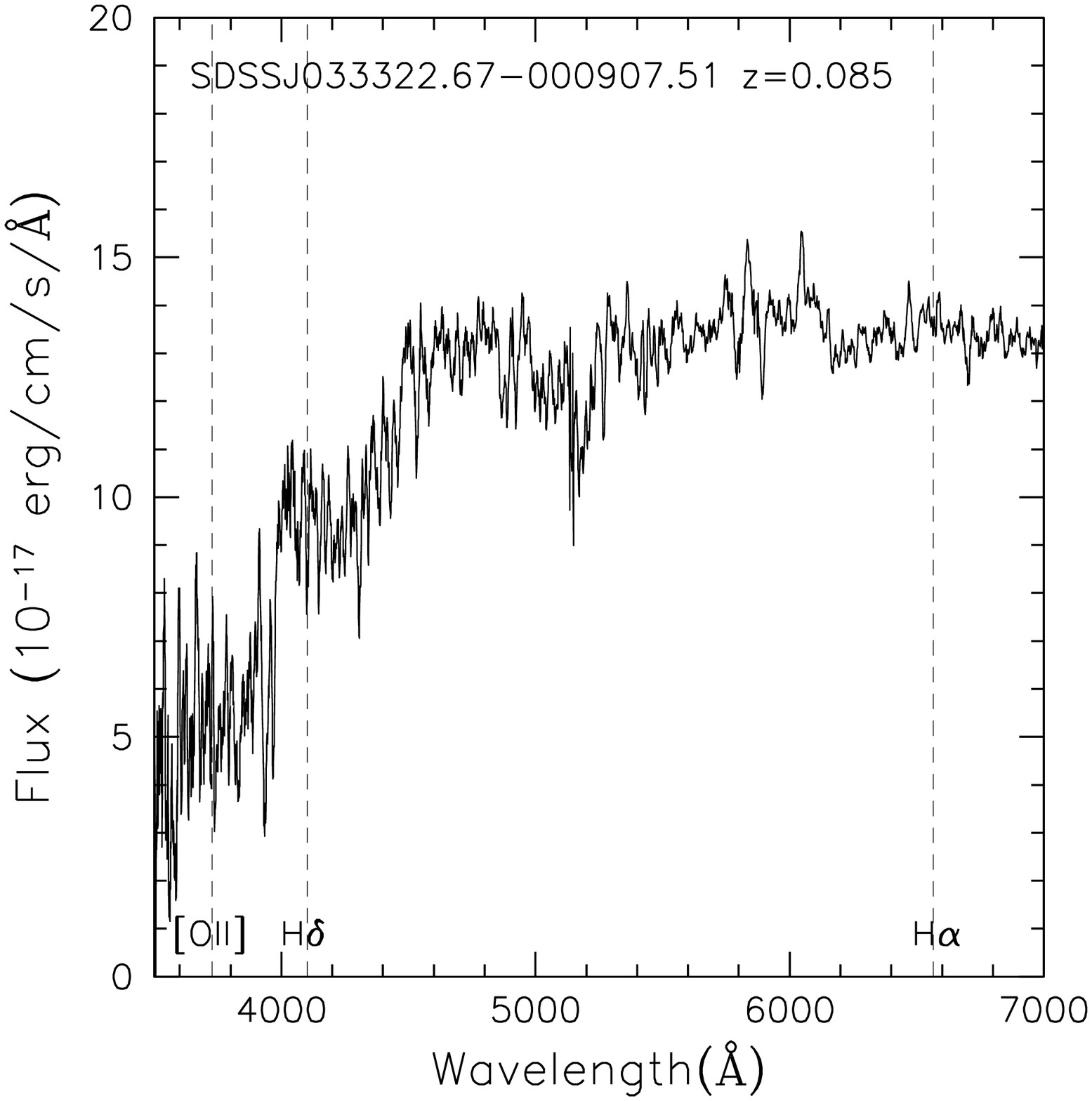}
\caption{(Left) Nine images of ``Passive'' spiral galaxies found by
Goto et al. (2003) in their search for anemic galaxies in the SDSS
database. Note the lack of ``blue'' HII star--forming regions in the
arms of these spirals. (Right) The SDSS spectrum for one of these
``Passive Spirals'' (the galaxy in the top left--hand corner). Notice
the lack of any emission lines indicative of on--going
star--formation. This plot was taken from Goto et al. (2003).}
\label{pass}
\end{figure}

\section{Strangulation of Star--Formation}

One possible physical model$\footnote{See the reviews of Bower, Mihos
\& Moore for a discussion of other physical mechanisms that can affect
the properties of galaxies in clusters and groups of galaxies}$ for
explaining the critical density seen at $\simeq 1 h^2_{75}{\rm
Mpc^{-2}}$ (or $>1$ virial radius) in the SFR--Density and \mr
relations is the stripping of the warm gas in the outer halo of 
in--falling spiral galaxies, via tidal interactions with the cluster
potential. This process removes the reservoir of hydrogen which
replenishes the gas in the cold disk of the galaxy, and thus slowly
strangles (or starves) the star--formation in the disk leading to a
slow death (see Larson et al. 1980; Diaferio et al. 2000; Balogh et
al. 2000). Recent N--body simulations of this model by Bekki et
al. (2002) demonstrate that it is a viable and can happen at large
cluster radii (or low densities).

One possible observational consequence of this strangulation of
star--formation is the existence of red, or passive, spiral galaxies,
{\it i.e.}, galaxies that possess a spiral morphology, but have no
observed on--going star--formation. Such galaxies have been found
already in studies of high redshift clusters of galaxies (Couch et
al. 1998) and have been known for sometime at low redshift as ``Anemic
Spirals'' (see van den Bergh 1991), although their true nature has
been debated by many (see Bothun \& Sullivan 1980; Guiderdoni 1987 and
the review of van Gorkom in these proceedings). Goto et al. (2003) has
performed an automated searched for such ``Passive'' or ``Anemic''
spirals within the SDSS database by looking for galaxies with a high
\cin value (indicating a face--on spiral galaxy; see Shimasaku et
al. 2001) but with no detected emission lines in their SDSS
spectra. In total, they found 73 such galaxies, which comprises of
only 0.28$\pm$0.03\% of all spiral galaxies with the same \cin
parameter values, but with detected emission lines. I show some
examples of these ``Passive Spirals'' in Figure \ref{pass}.

The most interesting discovery of Goto et al. is the distribution of
local densities for these ``Passive Spirals'', which peaks at $\simeq
1 h^2_{75}{\rm Mpc^{-2}}$. Therefore, these galaxies appear to be
preferential located close to the critical density discussed above for
the SFR--Density and \mr relations, {\it i.e.}, in the in--fall
regions of C4 clusters. This is consistent with the observed decrease
in the late spiral galaxies (Sc's) witnessed in Figure \ref{critical}.

In summary, the quality and quantity of the SDSS data allows us to
study the environmental dependences of galaxy properties in greater
detail than before.  There does appear to be at least one critical
density (at $\simeq 1 h^2_{75}{\rm Mpc^{-2}}$) affecting the
properties of galaxies and this could be due to the slow strangulation
of star--formation in these spiral galaxies as they fall into denser
environments.


I would like to acknowledge the organizers of the conference for their
invitation to participate and their hospitality in Pasadena. I also
thank all my collaborators for allowing me to show their results and
figures in this review. These include Chris Miller, Daniel Eisenstein,
Risa Wechsler, Gus Evrard, Tim McKay, Tomo Goto, Michael Balogh, Ann
Zabuldoff, and my colleagues from both the SDSS--2dF LRG survey and
SDSS. I thank Kathy Romer and Chris Miller for reading an earlier
draft of this review.

Funding for the creation and distribution of the SDSS Archive has been
provided by the Alfred P. Sloan Foundation, the Participating
Institutions, the National Aeronautics and Space Administration, the
National Science Foundation, the U.S. Department of Energy, the
Japanese Monbukagakusho, and the Max Planck Society. The SDSS Web site
is http://www.sdss.org/.

The SDSS is managed by the Astrophysical Research Consortium (ARC) for
the Participating Institutions. The Participating Institutions are The
University of Chicago, Fermilab, the Institute for Advanced Study, the
Japan Participation Group, The Johns Hopkins University, Los Alamos
National Laboratory, the Max-Planck-Institute for Astronomy (MPIA),
the Max-Planck-Institute for Astrophysics (MPA), New Mexico State
University, University of Pittsburgh, Princeton University, the United
States Naval Observatory, and the University of Washington.


\begin{thereferences}{}


\bibitem{bahcalk} Bahcall, N. A., et al., 2003, \apj, submitted. 

\bibitem{2000ApJ...540..113B} Balogh, 
M.~L., Navarro, J.~F., \& Morris, S.~L.\ 2000, \apj, 540, 113 


\bibitem{1990AJ....100...32B} Beers, 
T.~C., Flynn, K., \& Gebhardt, K.\ 1990, \aj, 100, 32 



\bibitem{2002astro.ph..6207B} Bekki, K., 
Shioya, Y., \& Couch, W.~J.\ 2002, ArXiv Astrophysics e-prints, 6207 




\bibitem{2000MNRAS.316..107B} Benson, A.~J., Baugh, 
C.~M., Cole, S., Frenk, C.~S., \& Lacey, C.~G.\ 2000, \mnras, 316, 107 

\bibitem{2001AJ....121.2358B} Blanton, M.~R.~et al.\ 
2001, \aj, 121, 2358 


\bibitem{2003} Blanton, M.R., Lupton, R.H., Maley, F.M., Young, N., Zehavi, I., Loveday, J. 2003, AJ, 125, 2276


\bibitem{2001A&A...369..826B} B{\" o}hringer, 
H.~et al.\ 2001, A\&A, 369, 826 

\bibitem{2000ApJ...533..601B} Bramel, D.~A., 
Nichol, R.~C., \& Pope, A.~C.\ 2000, \apj, 533, 601 


 \bibitem{1980ApJ...242..903B} Bothun, G.~D.~\& 
Sullivan, W.~T.\ 1980, \apj, 242, 903 



\bibitem{2000MNRAS.319..939C} Collins, C.~A.~et al.\ 
2000, \mnras, 319, 939 

\bibitem{1998ApJ...497..188C} Couch, W.~J., Barger, 
A.~J., Smail, I., Ellis, R.~S., \& Sharples, R.~M.\ 1998, \apj, 497, 188 



\bibitem{2001MNRAS.323..999D} Diaferio, A., 
Kauffmann, G., Balogh, M.~L., White, S.~D.~M., Schade, D., \& Ellingson, 
E.\ 2001, \mnras, 323, 999 


\bibitem{1980ApJ...236..351D} Dressler, A.\ 1980, \apj, 
236, 351 

\bibitem{1997ApJ...490..577D} Dressler, A.~et al.\ 
1997, \apj, 490, 577 


\bibitem{2001AJ....122.2267E} Eisenstein, D.~J.~et 
al.\ 2001, \aj, 122, 2267 

\bibitem{2003ApJ...586..718E} Eisenstein, D.~J.\ 2003, 
\apj, 586, 718 



\bibitem{2002ApJ...573....7E} Evrard, A.~E.~et al.\ 
2002, \apj, 573, 7 

\bibitem{1996AJ....111.1748F} Fukugita, M., 
Ichikawa, T., Gunn, J.~E., Doi, M., Shimasaku, K., \& Schneider, D.~P.\ 
1996, \aj, 111, 1748 



\bibitem{2000AJ....120.2148G} Gladders, M.~D.~\& 
Yee, H.~K.~C.\ 2000, \aj, 120, 2148 


\bibitem{2003ApJ...584..210G} G{\' o}mez, P.~L.~et 
al.\ 2003, \apj, 584, 210 


\bibitem{2002AJ....123.1807G} Goto, T.~et al.\ 2002, 
\aj, 123, 1807 

\bibitem{2003astro.ph..1303G} Goto, T.~et al.\ 2003, 
ArXiv Astrophysics e-prints, 1303 

\bibitem{1998AJ....116.3040G} Gunn, J.~E.~et al.\ 1998, 
\aj, 116, 3040 



\bibitem{1987A&A...172...27G} Guiderdoni, B.\ 1987, A\&A, 
172, 27 

\bibitem{2001AJ....122.2129H} 
Hogg, D.~W., Finkbeiner, D.~P., Schlegel, D.~J., \& Gunn, J.~E.\ 2001, \aj, 
122, 2129 


\bibitem{2003ApJ...585L...5H} Hogg, D.~W.~et al.\ 2003, 
\apjl, 585, L5 

\bibitem{2001MNRAS.321..372J} Jenkins, A., Frenk, 
C.~S., White, S.~D.~M., Colberg, J.~M., Cole, S., Evrard, A.~E., Couchman, 
H.~M.~P., \& Yoshida, N.\ 2001, \mnras, 321, 372 


\bibitem{1999ApJ...517...78K} Kepner, J., Fan, X., 
Bahcall, N., Gunn, J., Lupton, R., \& Xu, G.\ 1999, \apj, 517, 78 



\bibitem{2002AJ....123...20K} Kim, R.~S.~J.~et al.\ 2002, 
\aj, 123, 20 


\bibitem{2003ApJ...585..161K} Kochanek, C.~S., 
White, M., Huchra, J., Macri, L., Jarrett, T.~H., Schneider, S.~E., \& 
Mader, J.\ 2003, \apj, 585, 161 


\bibitem{1980ApJ...237..692L} Larson, 
R.~B., Tinsley, B.~M., \& Caldwell, C.~N.\ 1980, \apj, 237, 692 


\bibitem{Lee} Lee, B., et al., 2003, \apj, submitted. 


\bibitem{2002MNRAS.334..673L} Lewis, I.~et al.\ 2002, 
\mnras, 334, 673 



\bibitem{1983MNRAS.204...33L} Lucey, J.~R.\ 1983, \mnras, 204, 
33 

\bibitem{2001adass..10..269L} Lupton, R.~H., Gunn, 
J.~E., Ivezi{\' c}, Z., Knapp, G.~R., Kent, S., \& Yasuda, N.\ 2001, ASP 
Conf.~Ser.~238: Astronomical Data Analysis Software and Systems X, 10, 269 




\bibitem{2000PhDT........16M} Miller, C.~J.\ 2000, 
Ph.D.~Thesis,  

 \bibitem{2001AJ....122.3492M} Miller, C.~J.~et al.\ 
2001, \aj, 122, 3492 



\bibitem{1992MNRAS.255P..21N} 
Nichol, R.~C., Collins, C.~A., Guzzo, L., \& Lumsden, S.~L.\ 1992, \mnras, 
255, 21P 

\bibitem{2001misk.conf..613N} Nichol, R.~C.~et al.\ 
2001, Mining the Sky, 613 

\bibitem{2002tceg.conf...57N} Nichol, R.~C.\ 2002, ASP Conf.~Ser.~268:
Tracing Cosmic Evolution with Galaxy Clusters, 57

\bibitem{2002tceg} Pier, J.R., Munn, J.A., Hindsley, R.B., Hennessy, G.S., Kent, S.M., Lupton, R.H., and Ivezic, Z. 2003, AJ, 125, 1559 

\bibitem{1991MNRAS.253..307P} Peacock, J.~A.~\& 
Nicholson, D.\ 1991, \mnras, 253, 307 

\bibitem{1984ApJ...281...95P} Postman, M.~\& 
Geller, M.~J.\ 1984, \apj, 281, 95 


\bibitem{1992ApJ...384..404P} Postman, 
M., Huchra, J.~P., \& Geller, M.~J.\ 1992, \apj, 384, 404 

\bibitem{1996AJ....111..615P} Postman, M., Lubin, 
L.~M., Gunn, J.~E., Oke, J.~B., Hoessel, J.~G., Schneider, D.~P., \& 
Christensen, J.~A.\ 1996, \aj, 111, 615 


\bibitem{2002tceg.conf....3P} Postman, M.\ 2002, ASP 
Conf.~Ser.~268: Tracing Cosmic Evolution with Galaxy Clusters, 3 


\bibitem{1998ApJ...500..525S} 
Schlegel, D.~J., Finkbeiner, D.~P., \& Davis, M.\ 1998, \apj, 500, 525 

\bibitem{2001AJ....122.1238S} Shimasaku, K.~et al.\ 
2001, \aj, 122, 1238 



\bibitem{2002AJ....123..485S} Stoughton, C.~et al.\ 
2002, \aj, 123, 485 


\bibitem{2002AJ....124.1810S} Strauss, M.~A.~et al.\ 
2002, \aj, 124, 1810 


\bibitem{1988MNRAS.234..159S} Sutherland, W.\ 1988, 
\mnras, 234, 159 

\bibitem{2003astro.ph..3267T} Treu, T., Ellis, R.~S., 
Kneib, J.~-., Dressler, A., Smail, I., Czoske, O., Oemler, A., \& 
Natarajan, P.\ 2003, ArXiv Astrophysics e-prints, 3267 


\bibitem{1991PASP..103..390V} van den Burgh, S.\ 1991, 
\pasp, 103, 390 

\bibitem{2000AJ....120.1579Y} York, D.~G.~et al.\ 2000, 
\aj, 120, 1579 


\bibitem{2002ApJ...571..172Z} Zehavi, I.~et al.\ 2002, 
\apj, 571, 172 




























\end{thereferences}
\end{document}